\documentclass[12pt,fleqn]{article}
\usepackage{latexsym}

\begin{document}
\title{Time and Ensemble Averages in Bohmian Mechanics}
\author{Yakir Aharonov$^{a,b}$, Noam Erez$^{a}\footnote{Corresponding author. Present affiliation: (c). E-mail:noam@jewel.tamu.edu}${\ } and Marlan O. Scully$^{c,d}$
 {\ }\\
\small a) {\em \small School of Physics and Astronomy, Tel Aviv
University, Tel Aviv 69978, Israel}\\ \small b) {\em \small
Department of Physics, University of South Carolina, Columbia, SC
29208}\\ \small c) {\em \small Institute for Quantum Studies and
Department of Physics, Texas A\&M University,}\\ \small {College
Station, \small TX 77843-4242, USA} \\ \small d) {\em \small
Max-Planck-Institut f\"{u}r Quantenoptik,}\\ {\small
Hans-Kopfermann-Strasse 1, 85748 Garching, Germany}
}

\maketitle
\begin{abstract} We show that in the framework of one-dimensional
Bohmian Quantum Mechanics\cite{Bohm}, for a particle subject to a potential
undergoing a weak adiabatic change, the time averages of the
particle's positions typically differ markedly from the ensemble
averages. We Apply this result to the case where the weak
perturbing potential is the back-action of a measuring device
(i.e. a protective measurement). It is shown that under these
conditions, most trajectories never cross the position measured
(as already shown for a particular example in \cite{original}).
\end{abstract}

\vspace{.1 in}

\section{Introduction}

In the literature it is often stated that in the Bohmian picture, what is measured in a single
measurement is the particle position, whereas the wavefunction can be measured only indirectly.
However, the measurement process is described by the same Von Neumann model in which the interaction of the pointer of the measurement device
with the measured particle is the result of an interaction term in the Hamiltonian governing the
dynamics of their common \emph{wavefunction}. It is well known that for wavefunctions which are eigenstates of certain Hamiltonians, 
the particle's Bohmian position is constant in time\footnote{The precise condition is that the probability density current, J, vanish identically for the eigenstates. 
This is typically the case with Hamiltonians that are invariant under time reversal, for nondegerate
eigenstates (the last requirement is also needed for the protective measurement in what follows). We will also restrict our attention to bounded states.}.
However, one might hope that the act of measurement itself, by introducing a perturbing potential,
somehow causes the time distribution of the position to resemble the ensemble distribution (which by hypothesis coincides with the 
wavefunction distribution). 

In classical statistical mechanics, the ergodic hypothesis tells us that
we have two ways of measuring the probability of a particle to be in a region -
to measure the appropriate density for many particles (averaging over a Boltzmann ensemble), or to
track a single particle over a long time and calculate the proportion of the time it spends there to
the total. 
In fact, for some quantum systems we can do something similar: we can either measure the probability to be in a region by measuring the projection operator onto the region, 
for a large number of identically prepared particles (the usual ensemble average), or we can gently measure this operator for a \emph{single} particle 
over a long time (protective measurement, which will be described below). Since the protective measurement\cite{PM} is gentle (weak and adiabatic), 
it hardly changes the wavefunction, and so any time averages are trivial. Could the Bohmian particle position play the part 
of the ``microscopic details'' of statistical mechanics?

 If indeed what one measures is the particle position, then a time average over this should, 
during a protective measurement, reproduce the ensemble probability distribution (since we know that that is in fact what is measured). 
We show here that, on the contrary, during such a measurement, for most initial positions of the particle, the particle never reaches the region in question, 
and so the average time it spends there is necessarily different than what is being measured. 

 The equivalence of the predictions of Bohmian and quantum mechanics requires only that ensemble probabilities
for the Bohmian position correspond to those defined by the wavefunction. That this condition can indeed be satisfied is the fundamental theorem of Bohmian mechanics.
This does not require any ergodic property. However, one should bear this distinction in mind when interpreting the result of a measurement: for a measurment on a single
particle, one cannot assume that the result is determined by the particle postion---not even when it is uniquely determined by the wavefunction\footnote{Some authors have suggested that the auxiliary condition of Bohmian mechanics, 
that at an initial time the ensemble distribution agrees with the quantum probability, might be reduced from a postulate to an equilibrium condition which is
approached by most ensembles for a general sytem. This hypothesis---the ``Bohmian H-Theorem''---is equivalent to a kind of ergodic postulate. Our result 
demonstrates that at least for a single particle moving in one dimension, this cannot in general be true.} 
(as in a protective measurement)!

In a protective measurement, we start with an eigenwavefunction, and apply a weak adiabatic interaction with a 
``pointer'', which acts for a long time. It has been shown\cite{AAV} that (in the adiabatic limit), the wavefunction is unchanged by the measurement,
and it can therefore be measured (one region at a time) for a \emph{single} particle.
It was shown already in \cite{original}, for a particular Hamiltonian, that the ergodic hypothesis fails for this kind of measurment 
Here we show that this is true for a very broad class of potentials.

Similar problems with the realistic interpretation of the particle trajectory 
were discussed in the context of which-way measurements\cite{SES} in \cite{ESSW}, and in the context of weak measurements in \cite{AV}.
A selection of responses to these papers from a Bohmian quantum mechanics perspective, is offered by Refs.\cite{Bohmian1,Bohmian2,Bohmian3,Bohmian4}.

In the next section we show that the stationary property of Bohmian particle trajectories when the wavefunctions are 
eigenfunctions of the Hamiltonian, has a simple generalization to the case where this Hamiltonian is changed adiabatically. In the last section we
discuss the implications for measurement theory.

\section{Bohmian trajectories in the adiabatic limit}

The principal feature of the Bohmian Interpretation of Quantum
Mechanics \cite{Bohm, Holland} that ensures the agreement of its predictions with those
of the standard one, is the fact that an ensemble of particles
distributed according to the wavefunction probability at an
initial moment will remain in agreement with the instantaneous
wavefunction at any subsequent moment. This essential property is,
in turn, ensured by the equation of motion of the particle's
position:

\begin{equation}
\dot{\vec{x}}=\vec{j}/\rho \ \label{eqmot}
\end{equation}where $\vec{x}(t)$ is the Bohmian particle position, $\rho(\vec{x} ) =
|\psi(\vec{x} )|^2$, and \newline $\vec{j}=\frac{\hbar}{2 i m} \{\psi^*\vec{\nabla}\psi-\psi\vec{\nabla}\psi^*\}$. The Newtonian equation of motion is
second order in $t$, and a complete description of a statistical
ensemble is given by the joint distribution of $\vec{x}$ and $\vec{p}$ (i.e. a
phase-space distribution). A Bohmian ensemble is completely
described by a distribution of $\vec{x}$ alone. Let us denote this
distribution by $\rho'(\vec{x},t)$. The Bohmian interpretation
postulates that $\rho'=\rho$ for some initial moment, and the
equations of motion insure that the equality remains true at later
times. Indeed, eq.(\ref{eqmot}) together with the usual
Schr\"{o}dinger equation for $\psi$ imply that $\rho$ and $\rho'$
are related by:

\begin{equation} \frac{\partial\rho'}{\partial t}= - \vec{\nabla}\cdot
\vec{j}
= \frac{\partial\rho}{\partial t}\label{cont}
\end{equation} where $\rho'$ is the ensemble distribution (which at time $t_0$
is postulated to coincide with $\rho=|\psi|^2$).

As noted above, it is well known that for a broad class of Hamiltonians, when the wavefunction is an energy eigenstate, the Bohmian particle position is constant.
One might expect that during a protective measurement (which is both weak and adiabatic) of a wavefunction which is initially in an
such an energy eigenstate, the situation will not be very different. This is borne out by the following analysis.

The equation of continuity (\ref{cont}) leads, for a one-dimensional
particle in a bound state, to the following result:
\vspace{.1 in}

\textbf{Lemma}: if at time $t_0$ the particle is located at $x_0$,
and $P_{\psi(t=t_0)}[x<x_0]=P_0$, at a subsequent time $t_1$, it
will be at $x_1$ such that $P_{\psi(t=t_1)}[x<x_1]=P_0$.
\vspace{.1 in}

\textbf{Proof}: This follows from the fact that the particle ensemble changes in such a way that it always remains in agreement with the wavefunction probability, and
the property that particle trajectories cannot cross (a consequence of the equation of motion).\newline
Formally, for the position of a particle obeying eq.(\ref{eqmot}), $x(t)$, we have:
\begin{eqnarray}
  \frac{\partial}{\partial t} P_{\psi(t)}[x<x(t)] & = & \frac{\partial}{\partial t}\int_{-\infty}^{x(t)} \rho(\xi,t)d \xi =
  \dot{x}(t)\rho(x(t),t)+\int_{-\infty}^{x(t)}\dot{\rho}(\xi,t)d\xi = \nonumber \\ \dot{x}\rho - \int_{-\infty}^{x}\frac{\partial j}{\partial x}d\xi & = &
  \dot{x}(t)\rho(x(t),t) - j(x(t),t) = 0.
\end{eqnarray}$\Box$

\vspace{.1 in} \section{Protective measurements of position}
Consider a Bohmian particle in a potential well, whose initial wavefunction
is the ground state. If we introduce an adiabatic and weak
perturbation of the potential which eventually goes to zero, we
know that the wavefunction coincides at any moment with the ground
state of the instantaneous Hamiltonian (we assume that the ground
state is always nondegenerate). Our assumptions about the
perturbation insure that the change in the wavefunction is small
at all times and eventually vanishes. The lemma then tells us that
the change in particle position is likewise small at all times and
vanishes for large times. If the potential is changed in a small
spatial interval, then most initial positions for the particle
lead to trajectories which never reach the aforementioned
interval. So for a Bohmian particle in a given position, we can probe the
wavefunction in most other positions without the particle ever
being present there.

\begin{center}
ACKNOWLEDGEMENTS
\end{center}

It is our pleasure to thank Lev Vaidman for helpful discussions.
Y.A. and N.E. would like to acknowledge the support by the Basic Research Foundation of
the Israeli Academy of Sciences and Humanities and by the National Science Foundation. M.O.S. acknowledges the
support of the Welch Foundation.

\end{document}